\documentstyle[12pt]{article}

\begin{document}
\begin{center}
{\bf \large On the Hopf structure of
$\mbox{\boldmath $U_{p,q}(gl(1|1))$}$  and
the universal $\mbox{\boldmath ${\cal T}$}$-matrix of
$\mbox{\boldmath $Fun_{p,q}(GL(1|1))$}$ }

\vspace{.75cm}

{\bf R. Chakrabarti}$^1${\footnote{E-mail: ranabir@unimad.ernet.in}},
{\bf R. Jagannathan}$^{2,3\,}${\footnote{E-mail: jagan@imsc.ernet.in}}

\medskip

$^1${\it Department of Theoretical Physics, University of Madras, \\
Guindy Campus, Madras - 600025, INDIA} \\
$^2${\it International Centre for Theoretical Physics,
34100 Trieste, ITALY} \\
$^{3\,\,}${\footnote{Permanent address}}{\it The Institute of
Mathematical Sciences \\
C.I.T Campus, Tharamani, Madras - 600113, INDIA}

\end{center}

\baselineskip18pt

\vspace{2cm}

\noindent {\bf Abstract}

Using the technique developed by Fronsdal and Galindo (Lett. Math.
Phys. 27 (1993) 57) for studying the Hopf duality between the
quantum algebras $Fun_{p,q}(GL(2))$ and $U_{p,q}(gl(2))$, the Hopf
structure of $U_{p,q}(gl(1|1))$, dual to $Fun_{p,q}(GL(1|1))$, is
derived and the corresponding universal ${\cal T}$-matrix of
$Fun_{p,q}(GL(1|1))$, embodying the suitably modified exponential
relationship $U_{p,q}(gl(1|1))$ $\rightarrow$ $Fun_{p,q}(GL(1|1))$,
is obtained.

\newpage

\noindent {\bf 1. Introduction}

\bigskip

Recently Fronsdal and Galindo [1] have studied the
duality structure relating $Fun_{p,q}(GL(2))$ and $U_{p,q}(gl(2))$
(see also [2,3] for some aspects of a similar approach to
$Fun_q(GL(2))$ and $Fun_q(SL(2))$ ) wherein
the transfer matrix $T$ gets replaced by the
representation-independent universal ${\cal T}$-matrix.
The universal ${\cal T}$-matrix, identified with the dual form

\renewcommand{\theequation}{1.{\arabic{equation}}}
\setcounter{equation}{0}

\begin{equation}
{\cal T} = \sum_A\,x^A \otimes X_A\,,
\end{equation}

\noindent where $\left\{ x^A \right\}$ and $\left\{ X_A \right\}$
are respectively the basis elements of a pair of dually conjugate
Hopf algebras ${\cal A}$ and ${\cal U}$, expresses the generalization
of the familiar exponential mapping ($g$ $\rightarrow$ $G$) in the
case of Lie algebras.  In the case of $Fun_{p,q}(GL(2))$ the
corresponding universal ${\cal T}$-matrix has been explicitly shown
[1] to have the all the required algebraic properties [4] of the
$T$-matrix.  Further, in the universal ${\cal T}$-matrix one has all
the representations of a quantum matrix pseudo-group
$Fun_{\ldots}(G)$ expressed in
terms of the matrix representations of its dually paired (in the
Hopf sense) quantized universal enveloping algebra $U_{\ldots}(g)$
(see [2] for the case of $Fun_{q}(GL(2))$).
Apart from considering the universal ${\cal T}$-matrix of
$Fun_q(SL(2))$ in detail, following [1], the universal
${\cal T}$-matrices of some inhomogeneous quantum groups
have also been obtained in [5] using suitable contraction techniques.
With the aim of providing a new explicit example of the universal
${\cal T}$-matrix we consider here the two-parametric($p,q$) Hopf
superalgebras corresponding to the classical $GL(1|1)$ and $gl(1|1)$.
We study the duality structure relating $Fun_{p,q}(GL(1|1))$ [6,7] and
$U_{p,q}(gl(1|1))$ [7], following [1], and construct the universal
${\cal T}$-matrix of $Fun_{p,q}(GL(1|1))$.  When $p$ $=$ $q$ the
analysis will reduce to the single-parameter case (in this context,
see [8,9] for some recent studies on the finite-dimensional irreducible
representations of $U_q(gl(n|m))$).

\vspace{1cm}

\noindent {\bf 2. Hopf structure of
$\mbox{\boldmath $Fun_{p,q}(GL(1|1))$}$}

\bigskip

\noindent The algebra $Fun_{p,q}(GL(1|1))$ is generated
by the elements of the $2 \times 2$ matrix

\renewcommand{\theequation}{2.{\arabic{equation}}}
\setcounter{equation}{0}

\begin{equation}
T = \left(
\begin{array}{cc}
a & \beta \\
\gamma & d
\end{array}
\right)\,,
\end{equation}

\noindent obeying the braiding relations

\begin{eqnarray}
a \beta & = & p^{-1} \beta a\,, \quad
a \gamma = q^{-1} \gamma a\,, \quad
d \beta = p^{-1} \beta d\,, \quad
d \gamma = q^{-1} \gamma d\,, \nonumber \\
p^{-1} \beta \gamma & + & q^{-1} \gamma \beta = 0\,, \quad
ad - da = \left( q - p^{-1} \right) \beta \gamma \,, \nonumber \\
\beta^2 & = & 0\,, \quad
\gamma^2 = 0\,.
\end{eqnarray}

\noindent The even elements $a$ and $d$ are invertible.
The coproduct($\Delta$), counit($\epsilon$) and the
antipode($S$) maps are given respectively by

\begin{eqnarray}
\Delta (T) & = & T \dot{\otimes} T\,, \nonumber \\
\Delta \left( a^{-1} \right) & = & a^{-1} \otimes a^{-1} -
a^{-1} \beta a^{-1} \otimes a^{-1} \gamma a^{-1}\,, \nonumber \\
\Delta \left( d^{-1} \right) & = & d^{-1} \otimes d^{-1} -
d^{-1} \gamma d^{-1} \otimes d^{-1} \beta d^{-1}\,,
\end{eqnarray}

\begin{equation}
\epsilon (T) = 1\!{\rm l}\,,
\end{equation}

\begin{equation}
S(T) = T^{-1}\,, \quad
S \left( a^{-1} \right) = a - \beta d^{-1} \gamma \,, \quad
S \left( d^{-1} \right) = d - \gamma a^{-1} \beta \,,
\end{equation}

\noindent where

\begin{equation}
T^{-1} = \left(
\begin{array}{cc}
a^{-1} + a^{-1} \beta d^{-1} \gamma a^{-1} &
-a^{-1} \beta d^{-1} \\
-d^{-1} \gamma a^{-1} &
d^{-1} + d^{-1} \gamma a^{-1} \beta d^{-1}
\end{array}
\right)
\end{equation}

\noindent and $\dot{\otimes}$ denotes the tensor product combined
with the matrix multiplication.  Here, and throughout the paper,
the tensor product ($\otimes$) is a graded one corresponding to
superalgebras such that

\begin{equation}
\left( \Gamma_1 \otimes \Gamma_2 \right) \left( \Gamma_3 \otimes
\Gamma_4 \right) = (-1)^{{\rm deg} \left( \Gamma_2 \right) {\rm deg}
\left( \Gamma_3 \right)} \Gamma_1 \Gamma_3 \otimes
\Gamma_2 \Gamma_4\,,
\end{equation}

\noindent where ${\rm deg}( \Gamma )\,=\,0\,(1)$ for an even\,(odd)
$\Gamma$\,.  The invertible quantum superdeterminant

\begin{equation}
D = a d^{-1} - \beta d^{-1} \gamma d^{-1}
\end{equation}

\noindent is a central element of the algebra and follows a grouplike
coproduct rule

\begin{equation}
\Delta (D) = D \otimes D\,.
\end{equation}

\noindent The counit and antipode maps read

\begin{equation}
\epsilon (D) = 1\,, \quad
S(D) = D^{-1}\,,
\end{equation}

\noindent where

\begin{equation}
D^{-1} = d a^{-1} + \beta a^{_1} \gamma a^{_1}\,.
\end{equation}

A factorization of the $T$-matrix (2.1) as

\begin{equation}
T = \left( \begin{array}{cc}
           1 & 0 \\
           \zeta & 1
           \end{array} \right)
    \left( \begin{array}{cc}
           a & 0 \\
           0 & \hat{d}
           \end{array} \right)
    \left( \begin{array}{cc}
           1 & \xi \\
           0 & 1
           \end{array} \right)
\end{equation}

\noindent introduces new variables related to the old ones by

\begin{equation}
\beta = a \xi \,, \quad
\gamma = \zeta a\,, \quad
d = \zeta a \xi + \hat{d}\,,
\end{equation}

\noindent where $\hat{d}$ is an invertible element.  The quantum
superdeterminant and its inverse now read

\begin{equation}
D = a \hat{d}^{-1}\,, \quad
D^{-1} = \hat{d} a^{-1}\,.
\end{equation}

\noindent The algebra (2.2) assumes the form

\begin{eqnarray}
a \xi & = & p^{-1} \xi a\,, \quad
a \zeta = q^{-1} \zeta a\,, \quad
\hat{d} \xi = p^{-1} \xi \hat{d}\,, \quad
\hat{d} \zeta = q^{-1} \zeta \hat{d}\,, \nonumber \\
\left\{ \xi , \zeta \right\} & = & 0\,, \quad
\left[ a , \hat{d} \right] = 0\,, \quad
\xi^2 = 0\,, \quad
\zeta^2 = 0\,.
\end{eqnarray}

\noindent The coalgebra maps are rewritten as

\begin{eqnarray}
\Delta (a) & = & (a \otimes 1\!{\rm l})(1\!{\rm l} \otimes 1\!{\rm l}
+ \xi \otimes \zeta )(1\!{\rm l} \otimes a))\,, \nonumber \\
\Delta (\xi ) & = & 1\!{\rm l} \otimes \xi + \xi \otimes D^{-1}\,,
\quad
\Delta (\zeta ) = \zeta \otimes 1\!{\rm l} + D^{-1} \otimes \zeta \,,
\nonumber \\
\Delta \left( \hat{d} \right) & = & \left( \hat{d} \otimes 1\!{\rm l}
\right) (1\!{\rm l} \otimes 1\!{\rm l} + \xi \otimes \zeta )
\left( 1\!{\rm l} \otimes \hat{d} \right)\,,
\end{eqnarray}

\begin{equation}
\epsilon (a) = 1\,, \quad
\epsilon ( \xi ) = 0\,, \quad
\epsilon ( \zeta ) = 0\,, \quad
\epsilon \left( \hat{d} \right) = 1\,,
\end{equation}

\begin{eqnarray}
S(a) & = & a^{-1} + \xi a^{-1} D \zeta \,, \quad
S( \xi ) = - \xi D\,, \nonumber \\
S( \zeta ) & = & -D \zeta \,, \quad
S \left( \hat{d} \right) = \hat{d}^{-1} + \xi d^{-1} D \zeta \,.
\end{eqnarray}

\noindent A further map of the even generators

\begin{equation}
a = {\rm e}^x\,, \qquad
\hat{d} = {\rm e}^{\hat{x}}
\end{equation}

\noindent and a reparametrization

\begin{equation}
p = {\rm e}^{-\omega}\,\,, \qquad
q = {\rm e}^{-\nu}\,,
\end{equation}

\noindent give the algebra (2.15) a Lie superstructure

\begin{eqnarray}
\left[ x , \xi \right] & = & \omega \xi \,, \quad
\left[ x , \zeta \right] = \nu \zeta \,, \quad
\left[ \hat{x} , \xi \right] = \omega \xi \,, \quad
\left[ \hat{x} , \zeta \right] = \nu \zeta \,, \nonumber \\
\left[ x , \hat{x} \right] & = & 0\,, \quad
\left\{ \xi , \zeta \right\} = 0\,, \quad
\xi^2 = 0\,, \quad \zeta^2 = 0\,,
\end{eqnarray}

\noindent with the nontrivial coalgebra maps

\begin{eqnarray}
\Delta (x) & = & x \otimes 1\!{\rm l} + 1\!{\rm l} \otimes x
+ \Omega \xi \otimes \zeta \,, \quad
\Delta \left( \hat{x} \right) = \hat{x} \otimes 1\!{\rm l} +
1\!{\rm l} \otimes \hat{x} + \Omega \xi \otimes \zeta \,,
\nonumber \\
\Delta ( \xi ) & = & 1\!{\rm l} \otimes \xi + \xi \otimes
{\rm e}^{\hat{x} - x}\,\,, \quad
\Delta ( \zeta ) = \zeta \otimes 1\!{\rm l} + {\rm e}^{\hat{x} - x}
\otimes \zeta \,, \nonumber \\
   &   &  \quad \quad {\rm with} \ \ \ \Omega = \frac{\nu + \omega}
{{\rm e}^\nu - {\rm e}^{-\omega}}\,,
\end{eqnarray}

\begin{equation}
\epsilon ( x ) = \epsilon \left( \hat{x} \right)
= \epsilon ( \xi ) = \epsilon ( \zeta ) = 0\,,
\end{equation}

\begin{eqnarray}
S ( x ) & = & -x + \Omega \xi {\rm e}^{x - \hat{x}} \zeta \,, \quad
S \left( \hat{x} \right) = - \hat{x} + \Omega \xi
{\rm e}^{x - \hat{x}}\,\,, \nonumber \\
S ( \xi ) & = & - \xi {\rm e}^{x - \hat{x}}\,\,, \quad
S ( \zeta ) = - {\rm e}^{x - \hat{x}}\,\zeta \,.
\end{eqnarray}

It is now evident that $Fun_{p,q}(GL(1|1))$ may be embedded in the
enveloping algebra of a Lie superalgebra with a noncocommutative
coproduct structure.  Next, we show that this enveloping algebra is
dual to the Hopf superalgebra $U_{p,q}(gl(1|1))$.

\vspace{1cm}

\noindent {\bf 3. $\mbox{\boldmath
$Fun_{p,q}(GL(1|1))-U_{p,q}(gl(1|1))$}$ duality and the universal
$\mbox{\boldmath ${\cal T}$}$-matrix of
$\mbox{\boldmath $Fun_{p,q}(GL(1|1))$}$}

\bigskip

\noindent  Two Hopf algebras ${\cal A}$ and ${\cal U}$ are in
duality (see, e.g., [10] for details) if there exists a doubly
nondegenerate bilinear form

\renewcommand{\theequation}{3.{\arabic{equation}}}
\setcounter{equation}{0}

\begin{equation}
\langle \,,\,\rangle : (a,u) \rightarrow \langle a,u \rangle
\quad \forall \,a \in {\cal A}\,,\quad
\forall \,u \in {\cal U}\,,
\end{equation}

\noindent such that, for $(a,b) \in {\cal A}$,
$(u,v) \in {\cal U}$,

\begin{equation}
\langle a , uv \rangle = \left\langle \Delta_{\cal A}(a) , u
\otimes v \right\rangle \,, \quad
\langle ab , u \rangle = \left\langle a \otimes b ,
\Delta_{\cal U}(u) \right\rangle \,,
\end{equation}

\begin{equation}
\left\langle a , 1\!{\rm l}_{\cal U} \right\rangle =
\epsilon_{\cal A}(a)\,, \qquad
\left\langle 1\!{\rm l}_{\cal A} , u \right\rangle =
\epsilon_{\cal U}(u)\,,
\end{equation}

\begin{equation}
\left\langle a , S_{\cal U}(u) \right\rangle =
\left\langle S_{\cal A}(a) , u \right\rangle \,.
\end{equation}

\noindent Let $\left\{ e^A \mid e^A = \zeta^{a_1}
x^{a_2} \hat{x}^{a_3} \xi^{a_4}\,,\right.$ $A = \left( a_1 ,
\right.$$a_2 ,$
$a_3 ,$ $\left. a_4 \right)$ , $\left( a_1 , a_4 \right)$ $=$
$(0,1)$ , $\left(
a_2 , a_3 \right)$ $\left. \in Z\!\!\!Z_+ \right\}$ be a basis of
monomials for $Fun_{p,q}(GL(1|1))$ obeying the multiplication and
the induced coproduct rules

\begin{equation}
e^A e^B = \sum_{C}\,f^{AB}_C\,e^C\,,
\end{equation}

\begin{equation}
\Delta \left( e^A \right) = \sum_{BC}\,h^A_{BC}\,e^B
\otimes e^C\,.
\end{equation}

\noindent The unit element is obtained by choosing $A$ $=$
$\underline{0}$, where $\underline{0}$ $=$ $(0,0,0,0)$.  The dual
basis elements $\left\{ E_A \right\}$ are defined by

\begin{equation}
\left\langle e^A , E_B \right\rangle = \delta^A_B\,, \quad
\delta^A_B = \Pi_{i=1}^4\,\delta^{a_i}_{b_i}\,.
\end{equation}

\noindent Then, we obtain the following multiplication and
coproduct structures for the basis set $\left\{ E_A \right\}$:

\begin{equation}
E_A E_B = \sum_C\,h^C_{AB}\,E_C\,,
\end{equation}

\begin{equation}
\Delta \left( E_A \right) = \sum_{BC}\,f^{BC}_A\,E_B
\otimes E_C\,.
\end{equation}

Using the algebra (2.2) the structure tensor $f^{AB}_C$
is obtained:

\begin{eqnarray}
f^{AB}_C & = & (-1)^{a_4 b_1} \bar{\delta}^{a_1 b_1}
\bar{\delta}^{a_4 b_4} \delta^{a_1+b_1}_{c_1}
\theta^{a_2+b_2}_{c_2} \theta^{a_3+b_3}_{c_3}
\delta^{a_4+b_4}_{c_4} \nonumber \\
   &  & \sum_{kl}\,\left( \begin{array}{c}
                          a_2 \\
                          k \end{array} \right)
                   \left( \begin{array}{c}
                          b_2 \\
                          c_2 - k \end{array} \right)
                   \left( \begin{array}{c}
                          a_3 \\
                          l \end{array} \right)
                   \left( \begin{array}{c}
                          b_3 \\
                          c_3 - l \end{array} \right)
\left( \nu b_1 \right)^{a_2+a_3-k-l} \nonumber \\
   &   & \left( -\omega a_4 \right)^{b_2+b_3-c_2-c_3+k+l}\,,
\end{eqnarray}

\noindent where $\bar{\delta}^{ab} = \delta^a_0 \delta^b_0 +
\delta^a_1 \delta^b_0 + \delta^a_0 \delta^b_1$ and $\theta^a_b$
$=$ $1$ $(0)$ if $a$ $\geq$ $b$ $(<b)$.  Some special cases
relevant for later use are:

\renewcommand{\theequation}{3.11{\alph{equation}}}
\setcounter{equation}{0}

\begin{eqnarray}
f^{AB}_{1000} & = & \left( \delta^{a_1}_0 \delta^{b_1}_1 +
\delta^{a_1}_1 \delta^{b_1}_0 \right) \delta^{b_2}_0 \delta^{b_3}_0
\delta^{a_4}_0 \delta^{b_4}_0 \left( \nu b_1 \right)^{a_2+a_3}\,, \\
f^{AB}_{0001} & = & \delta^{a_1}_0 \delta^{b_1}_0 \delta^{a_2}_0
\delta^{a_3}_0 \left( \delta^{a_4}_1 \delta^{b_4}_0 + \delta^{a_4}_0
\delta^{b_4}_1 \right) \left( - \omega a_4 \right)^{b_2+b_3}\,, \\
f^{AB}_{0100} & = & \delta^{a_1}_0 \delta^{a_2}_1 \delta^{a_3}_0
\delta^{a_4}_0 \delta^B_{\underline{0}} + \delta^A_{\underline{0}}
\delta^{b_1}_0 \delta^{b_2}_1 \delta^{b_3}_0 \delta^{b_4}_0\,, \\
f^{AB}_{0010} & = & \delta^{a_1}_0 \delta{a_2}_0 \delta{a_3}_1
\delta{a_4}_0 \delta^B_{\underline{0}} + \delta^A_{\underline{0}}
\delta^{b_1}_0 \delta^{b_2}_0 \delta^{b_3}_1 \delta^{b_4}_0\,.
\end{eqnarray}

The tensor $h^A_{BC}$ is determined by the induced coproduct
rule for the basis set $\left\{ e^A \right\}$

\renewcommand{\theequation}{3.{\arabic{equation}}}
\setcounter{equation}{11}

\begin{equation}
\Delta \left( e^A \right) = \Delta ( \zeta )^{a_1}
\Delta (x)^{a_2} \Delta \left( \hat{x} \right)^{a_3}
\Delta ( \xi )^{a_4}\,.
\end{equation}

\noindent The following special cases, necessary for determining the
dual algebraic structure, may be directly read from (3.12):

\renewcommand{\theequation}{3.13{\alph{equation}}}
\setcounter{equation}{0}

\begin{eqnarray}
h^A_{B \underline{0}} & = & \delta^A_B\,, \qquad
h^A_{\underline{0} B} = \delta^A_B\,, \\
h^A_{0b_2b_300c_2c_30} & = & \delta^{a_1}_0 \delta^{a_2}_{b_2+c_2}
\delta^{a_3}_{b_3+c_3} \delta^{a_4}_0
\left( \begin{array}{c}
       a_2 \\
       b_2 \end{array} \right)
\left( \begin{array}{c}
       a_3 \\
       b_3 \end{array} \right)\,, \\
h^A_{1000B} & = & \delta^{a_1}_1 \delta^0_{b_1} \Pi_{i=2}
^4\,\delta^{a_i}_{b_i}i\,, \quad
h^A_{B0001} = \left( \Pi_{i=1}^3\,\delta^{a_i}_{b_i} \right)
\delta^{a_4}_1 \delta^0_{b_4}\,, \\
h^A_{01001000} & = & \delta^{a_1}_1 \delta^{a_2}_1 \delta^{a_3}_0
\delta^{a_4}_0 - \delta^{a_1}_1 \Pi_{i=2}^4\,\delta^{a_i}_0\,, \\
h^A_{00101000} & = & \delta^{a_1}_1 \delta^{a_2}_0 \delta^{a_3}_1
\delta^{a_4}_0 + \delta^{a_1}_1 \Pi_{i=2}^4\,\delta^{a_i}_0\,, \\
h^A_{00010100} & = & \delta^{a_1}_0 \delta^{a_2}_1 \delta^{a_3}_0
\delta^{a_4}_1 - \left( \Pi_{i=1}^3\,\delta^{a_i}_0 \right)
\delta^{a_4}_1i\,, \\
h^A_{00010010} & = & \delta^{a_1}_0 \delta^{a_2}_0 \delta^{a_3}_1
\delta^{a_4}_1 + \left( \Pi_{i=1}^3\,\delta^{a_i}_0 \right)
\delta^{a_4}_1\, \\
h^A_{00011000} & = & - \delta^{a_1}_1 \delta^{a_2}_0 \delta^{a_3}_0
\delta^{a_4}_1 + \delta^{a_1}_0 \delta^{a_4}_0 \sum_{n=1}^\infty
\,\Omega_n\,\delta^{a_2+a_3}_n\,,
\end{eqnarray}

\noindent where $\Omega_n = \frac{\nu^n - (-\omega )^n}
{{\rm e}^\nu - {\rm e}^{-\omega}}$.  The counit and the antipode
maps for the basis elements $\left\{ e^A \right\}$ are obtained
from (2.23) and (2.24) respectively:

\renewcommand{\theequation}{3.{\arabic{equation}}}
\setcounter{equation}{13}

\begin{equation}
\epsilon \left( e^A \right) = \delta^A_{\underline{0}}\,,
\end{equation}

\begin{eqnarray}
S \left( e^A \right) & = & (-1)^{a_1 a_4} S( \xi )^{a_4}
S \left( \hat{x} \right)^{a_3} S(x)^{a_2} S( \zeta )^{a_1}
\nonumber \\
  & = & (-1)^{\sum_{i=1}^4\,a_i+a_1a_4}\,\zeta^{a_1}
\left( \hat{x} + \nu a_1 - \omega a_4 + \Omega \zeta {\rm e}^
{x-\hat{x}} \xi \right)^{a_3} \nonumber \\
   &  &  \ \ \ {\rm e}^{ \left( a_1+a_4 \right)
         \left( x-\hat{x} \right)} \left( x + \nu a_1
         - \omega a_4 + \Omega \zeta {\rm e}^{x-\hat{x}}
         \xi \right)^{a_2} \xi^{a_4}\,.
\end{eqnarray}

\noindent The second equality in (3.15) is obtained by using the
commutation relations (2.21) and will be later used to compute the
antipode maps for the dual basis elements.

Employing the duality property, we now extract the multiplication
relations for the dual basis $\left\{ E_A \right\}$.  From (3.8) and
(3.13a) we obtain the unit element:

\begin{equation}
E_A E_{\underline{0}} = E_A\,, \quad
E_{\underline{0}} E_A = E_A\ \ \ \rightarrow \ \ \
E_{\underline{0}} = 1\!\!{\rm l}\,.
\end{equation}

\noindent The generators of the dual algebra are chosen as

\begin{equation}
E_- = E_{1000}\,, \quad H = E_{0100}\,, \quad
\tilde{H} = E_{0010}\,, \quad E_+ = E_{0001}\,.
\end{equation}

\noindent By repeated use of the relations (3.13b) and (3.13c)
we express an arbitrary dual basis element as

\begin{equation}
E_A = \left( a_2! a_3! \right)^{-1} E^{a_1}_- H^{a_2}
\tilde{H}^{a_3} E_+^{a_4}\,,
\end{equation}

\noindent where $\left( a_1 , a_4 \right)$ $=$ $(0,1)$ and
$\left( a_2 , a_3 \right)$ $\in$ $Z\!\!\!Z_+$.  Further use of the
special values of the structure tensor $h^A_{BC}$ in (3.13) now
yields the dual algebra $U_{p,q}(gl(1|1))$:

\begin{eqnarray}
\left[ H , E_\pm \right] & = & \pm E_\pm \,, \quad
\left[ \tilde{H} , E_\pm \right] = \mp E_\pm \,, \quad
\left[ H , \tilde{H} \right] = 0\,, \nonumber \\
\left\{ E_+ , E_- \right\} & = &
\frac{{\rm e}^{\nu \left( H + \tilde{H} \right)} -
{\rm e}^{-\omega \left( H + \tilde{H} \right)}}
{{\rm e}^\nu - {\rm e}^{-\omega }}\,, \quad
E^2_\pm = 0\,.
\end{eqnarray}

\noindent These relations, in turn, allow us to compute the
general expression for the structure tensor $h^A_{BC}$.  From
(3.8), (3.18) and (3.19), we get

\begin{eqnarray}
h^A_{BC} & = & (-1)^{b_2+c_2-a_2+c_1b_4} \bar{\delta}^{b_1c_1}
\bar{\delta}^{b_4c_4}
\delta^{b_1+c_1}_{a_1}
\theta^{b_2+c_2}_{a_2} \theta^{b_3+c_3}_{a_3}
\delta^{b_4+c_4}_{a_4} \nonumber \\
   &  & \ a_2!a_3! \left( b_2!b_3!c_2!c_3! \right)^{-1}
\sum_{kl}\,\left(
\begin{array}{c}
b_2 \\
k \end{array} \right) \left(
\begin{array}{c}
c_2 \\
a_2 - k \end{array} \right) \left(
\begin{array}{c}
b_3 \\
l \end{array} \right) \left(
\begin{array}{c}
c_3 \\
a_3 - l \end{array} \right) \nonumber \\
   &  & \ c_1^{b_2+b_3-k-l}
b_4^{c_2+c_3-a_2-a_3+k+l}
+ \delta^{b_1}_0 \delta^{c_1}_1 \delta^{a_1}_0
\delta^{b_4}_1 \delta^{c_4}_0 \delta^{a_4}_0
a_2!a_3! \nonumber \\
   &  & \ \left( b_2!c_2!b_3!c_3! \left( a_2-b_2-c_2 \right)!
\left( a_3-b_3-c_3 \right)! \right)^{-1} \nonumber \\
   &  & \ \sum_n\,\Omega_n
\delta^{a_2+a_3}_{b_2+b_3+c_2+c_3+n}\,.
\end{eqnarray}

The coproduct rules for the generators of the dual algebra
$U_{p,q}(gl(1|1))$ are obtained from (3.9) and (3.11):

\begin{eqnarray}
\Delta (H) & = & H \otimes 1\!{\rm l} + 1\!{\rm l} \otimes H\,,
\quad
\Delta \left( \tilde{H} \right) = \tilde{H} \otimes
1\!{\rm l} + 1\!{\rm l} \otimes \tilde{H}\,, \nonumber \\
\Delta \left( E_+ \right) & = & E_+ \otimes {\rm e}^{-\omega
\left( H + \tilde{H} \right)} + 1\!{\rm l} \otimes
E_+\,, \nonumber \\
\Delta \left( E_- \right) & = & E_- \otimes 1\!{\rm l} +
{\rm e}^{\nu \left( H + \tilde{H} \right)} \otimes E_-\,.
\end{eqnarray}

\noindent The counit maps for the dual generators are read from
(3.3) and (3.14):

\begin{equation}
\epsilon (X) = 0\,, \quad \forall X \in \left( H, \tilde{H},
E_\pm \right)\,.
\end{equation}

\noindent To determine the antipode maps for the dual generators,
we compute, using (3.7) and (3.15), the following elements of the
bilinear form:

\begin{eqnarray}
\left\langle S \left( e^A \right) , H \right\rangle
& = & - \delta^{a_1}_0 \delta^{a_2}_1 \delta^{a_3}_0
\delta^{a_4}_0\,, \quad
\left\langle S \left( e^A \right) , \tilde{H}
\right\rangle = - \delta^{a_1}_0 \delta^{a_2}_0
\delta^{a_3}_1 \delta^{a_4}_0\,, \nonumber \\
\left\langle S \left( e^A \right) , E_+ \right\rangle
& = & - \delta^{a_1}_0 \delta^{a_4}_1
\omega^{a_2+a_3}\,, \nonumber \\
\left\langle S \left( e^A \right) , E_- \right\rangle
& = & - \delta^{a_1}_1 \delta^{a_4}_0 (-\nu )^{a_2+a_3}\,.
\end{eqnarray}

\noindent The duality relation (3.4) now immediately yields the
antipode maps:

\begin{eqnarray}
S(H) & = & -H\,, \quad S \left( \tilde{H} \right) =
-\tilde{H}\,, \nonumber \\
S \left( E_+ \right) & = & -{\rm e}^{\omega \left( H+\tilde{H}
\right)} E_+\,, \quad
S \left( E_- \right) = -E_- {\rm e}^{-\nu \left( H+\tilde{H}
\right)}\,.
\end{eqnarray}

\noindent A map of the dual generators

\begin{equation}
Z = \frac{1}{2} \left( H+\tilde{H} \right)\,, \quad
J = \frac{1}{2} \left( H-\tilde{H} \right)\,, \quad
\chi_\pm = E_\pm Q^{\mp Z} \lambda^{-Z+\frac{1}{2}}\,,
\end{equation}

\noindent with $Q = \sqrt{pq}$ and $\lambda = \sqrt{p/q}$, now
reexpresses the Hopf structure of
$U_{p,q}(gl(1|1))$ ((3.19),(3.21),(3.22) and (3.24))
in the standard form:

\begin{eqnarray}
\left[ J , \chi_\pm \right] & = & \pm \chi_\pm \,, \quad
\left\{ \chi_+ , \chi_- \right\} = \frac{Q^{2Z}-Q^{-2Z}}
{Q-Q^{-1}}\,,
\quad \chi_\pm^2 = 0\,, \nonumber \\
\left[ Z , X \right] & = & 0\,, \ \ \ \
\forall X \in \left( J , \chi_\pm \right)\,,
\end{eqnarray}

\begin{eqnarray}
\Delta (Z) & = & Z \otimes 1\!{\rm l} + 1\!{\rm l} \otimes Z\,,
\quad
\Delta (J) = J \otimes 1\!{\rm l} + 1\!{\rm l} \otimes J\,,
\nonumber \\
\Delta \left( \chi_\pm \right) & = & \chi_\pm \otimes Q^Z
\lambda^{\pm Z} + Q^{-Z} \lambda^{\mp Z} \otimes \chi_\pm \,,
\end{eqnarray}

\begin{equation}
\epsilon (X) = 0\,, \quad
S(X) = -X\,, \quad \forall X \in \left( Z,J,\chi_\pm \right)\,.
\end{equation}

Finally, following the prescription (1.1) for the universal
${\cal T}$-matrix, we obtain the universal ${\cal T}$-matrix of
$Fun_{p,q}(GL(1|1))$ explicitly as

\begin{eqnarray}
{\cal T} & = & \sum_A\,e^A \otimes E_A \nonumber \\
  & = & \sum_{a_1,a_4 = 0}^1\,(-1)^{a_1a_4}\,\left( \zeta^{a_1}
\otimes E_-^{a_1} \right) {\rm e}^{x \otimes H}
{\rm e}^{\hat{x} \otimes \tilde{H}} \left( \xi^{a_4} \otimes
E_+^{a_4} \right)\,.
\end{eqnarray}

\noindent It is seen that corresponding to the two-dimensional
irreducible representation of the generators of
$U_{p,q}(gl(1|1))$ given by

\begin{equation}
E_+ = \left(
\begin{array}{cc}
0 & 1 \\
0 & 0
\end{array} \right)\,, \quad
E_- = \left(
\begin{array}{cc}
0 & 0 \\
1 & 0
\end{array} \right)\,, \quad
H = \left(
\begin{array}{cc}
1 & 0 \\
0 & 0
\end{array} \right)\,, \quad
\tilde{H} = \left(
\begin{array}{cc}
0 & 0 \\
0 & 1
\end{array} \right)\,,
\end{equation}

\noindent the above universal ${\cal T}$-matrix (3.29) reduces,
as required, to the $T$-matrix (2.1), read with (2.13) and (2.19):

\begin{equation}
T = \left(
\begin{array}{cc}
a & \beta \\
\gamma & d
\end{array} \right)
= \left(
\begin{array}{cc}
{\rm e}^x & {\rm e}^x \xi \\
\zeta {\rm e}^x & {\rm e}^{\hat{x}} + \zeta {\rm e}^x \xi
\end{array} \right)\,.
\end{equation}

\vspace{1cm}

\noindent{\bf 4. Conclusion}

\bigskip

To conclude, let us summarize.  Using the method developed by
Fronsdal and Galindo [1] for analysing the duality between the Hopf
algebras $Fun_{p,q}(GL(2))$ and $U_{p,q}(gl(2))$, we have extracted
the Hopf structure of the quantum superalgebra $U_{p,q}(gl(1|1))$
[7] from its duality relationship with $Fun_{p,q}(GL(1|1))$ [6,7]
obtained by a two-parametric$(p,q)$ quantization of the algebra of
functions on the supergroup $GL(1|1)$.  The
universal ${\cal T}$-matrix of $Fun_{p,q}(GL(1|1))$, identified
with the corresponding dual form, is seen to exhibit the suitably
modified exponential relationship $U_{p,q}(gl(1|1))$ $\rightarrow$
$Fun_{p,q}(GL(1|1))$.

\newpage

\noindent{\bf Acknowledgements}

One of us (R.J) thanks Prof. Abdus Salam,
the International Atomic Energy Agency and UNESCO for
hospitality at the International Centre for Theoretical Physics,
Trieste, where part of this work was done while visiting the
High Energy Section during August 1994; in this regard,
thanks are also due, in particular, to
Prof. S. Randjabar-Daemi and Prof. K.S. Narain.  He wishes also
to thank Prof. G. Baskaran for kind encouragement.  We dedicate
this paper to the memory of our beloved Prof. R. Vasudevan who
was to us much more than a senior colleague.

\end{document}